\documentclass{moriond}

\usepackage{graphicx}
\usepackage{amsmath}
\usepackage{multirow}
\usepackage{txfonts}
\newcommand{\Msol}{M_\odot}

\newcommand{\ud}{{\mathrm d}}

\def\be{\begin{equation}}
\def\ee{\end{equation}}
\def\bea{\begin{eqnarray}}
\def\eea{\end{eqnarray}}

\newcommand{\Photo}{}

\begin{document}
\vspace*{4cm}
\title{
Search for heavy blackholes with Microlensing: The MEMO project
}

\author{ M. Moniez, A. Mirhosseini}

\address{Laboratoire de l'Acc\'{e}l\'{e}rateur Lin\'{e}aire,
{\sc IN2P3-CNRS}, Universit\'e de Paris-Sud \\
B.P. 34, 91898 Orsay Cedex, France
}

\maketitle\abstracts{
The historical microlensing surveys MACHO, EROS, MOA and OGLE (hereafter summarized in
the MEMO acronym) have searched for microlensing toward the LMC for a total duration of 27 years.
We have studied the potential of joining all databases to search for very heavy objects
producing several year duration events.
We show that a combined systematic search for microlensing should detect of the order of 10 events due to $100\Msol$
black holes, that were not detectable by the individual surveys, if these objects have a
major contribution to the Milky-Way halo.
Assuming that a common analysis is feasible, {\it i.e.\rm} that the difficulties due to the use of different passbands
can be overcome, we show that the sensitivity of such an analysis should allow one to quantify the Galactic black hole component.
}

\section{Introduction}

Since 1989, several groups have operated survey programs
to search for compact halo objects within the Galactic halo,
following Paczy\'nskis' seminal publication \cite{pacz1986}.
The primitive challenge for the EROS (Exp\'erience de Recherche d'Objets Sombres)
and MACHO (MAssive Compact Halo Objects)
teams was to clarify the status of the missing hadrons in the Milky-Way.
Since the first discoveries by EROS \cite{eroslmc}, MACHO \cite{machlmc}, and
OGLE \cite{oglpr} (Optical Gravitational Lensing Experiment), thousands of microlensing effects have been
detected in the direction of the Galactic center together with
a handful of events toward the Galactic spiral arms \cite{BS7ans} and very
few events towards the Magellanic Clouds (LMC and SMC).
\\
Here, we focus on the data of the surveys towards the LMC \cite{MEMO},
the statistically dominant target to probe the dark compact objects of the Galactic halo (the others being the SMC and M31).
We explore the potential of a combined search for very long duration events due to heavy black holes that are now known to exist, thanks
to the recently discovered gravitational waves \cite{GW1}, and that can be
considered as possible candidates for the Galactic halo dark matter \cite{Bird}.

\section{Introduction to the microlensing effect}
\label{section:basics}
\begin{figure}[htbp]
\begin{center}
\includegraphics[width=9cm]{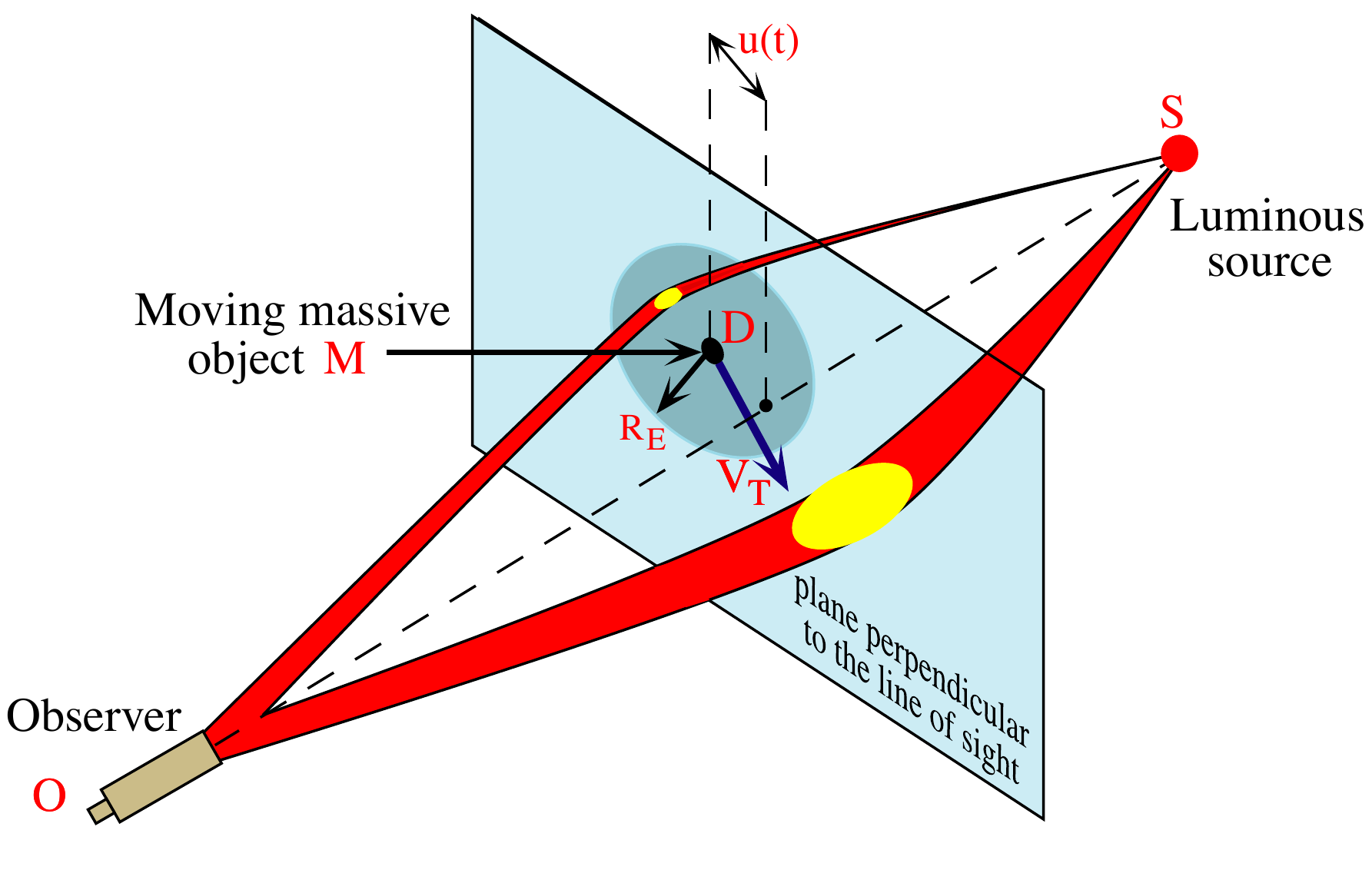}
\caption[]{
Principle of the microlensing effect: As the deflector D of mass M moves with a transverse relative velocity $V_T$, the impact parameter u(t) changes with time, and so does the magnification of the source
}
\label{principe}
\end{center}
\end{figure}

The gravitational microlensing effect is the temporary magnification of a source
when a massive compact object passes close enough to its line of sight (see Fig. \ref{principe})

Assuming a single point-like lens of mass $M$ located at distance $D_L$ is deflecting the
light from a single point-like source located at distance $D_S$, the magnification $A(t)$
of the source luminosity as a function of time $t$ is given by \cite{pacz1986} :
\begin{equation}
\label{magnification}
A(t)=\frac{u(t)^2+2}{u(t)\sqrt{u(t)^2+4}}\ ,
\end{equation}
where $u(t)$ is the distance of the lensing object to the undeflected line of sight, divided by
the Einstein radius $R_{\mathrm{E}}$ :
\begin{equation}
R_{\mathrm{E}} = \sqrt{\frac{4GM}{c^2}D_S x(1-x)}
\simeq 4.54\ \mathrm{AU}.\left[\frac{M}{\Msol}\right]^{\frac{1}{2}}
\left[\frac{D_S}{10 kpc}\right]^{\frac{1}{2}}
\frac{\left[x(1-x)\right]^{\frac{1}{2}}}{0.5},
\end{equation}
$G$ is the Newtonian gravitational constant, and $x = D_L/D_S$.
Assuming a lens moving at a constant relative transverse
velocity $v_T$, reaching its minimum
distance $u_0$ (impact parameter) to the undeflected line of sight
at time $t_0$, $u(t)$ is given by $u(t)=\sqrt{u_0^2+(t-t_0)^2/t_{\mathrm{E}}^2}$,
where $t_{\mathrm{E}}=R_{\mathrm{E}} /v_T$ is the lensing timescale:
\begin{eqnarray}
t_{\mathrm{E}} \sim
79\ \mathrm{days} \times 
\left[\frac{v_T}{100\, km/s}\right]^{-1}
\left[\frac{M}{\Msol}\right]^{\frac{1}{2}}
\left[\frac{D_S}{10\, kpc}\right]^{\frac{1}{2}}
\frac{[x(1-x)]^{\frac{1}{2}}}{0.5}\; . 
\end{eqnarray}

\subsection{Microlensing event characteristics}
The so-called simple microlensing effect (point-like source and lens
with rectilinear motions) has the following characteristic
features: 
The event results from a coincidence and is singular in the history of the source
(as well as of the deflector);
the magnification, independent of the color, is a simple function of time
depending only on ($u_0, t_0, t_{\mathrm{E}}$), symmetrical with respect to the time of maximum magnification;
as the source and the deflector are independent,
the prior distribution of the events' impact parameters must be uniform;
all stars at the same given distance have the same probability to be lensed;
therefore the sample of lensed stars should be representative
of the monitored population at that distance, particularly with respect to
the observed color and magnitude distributions.

This simple microlensing description can be complicated in
many different ways: for example, with multiple lens and source systems \cite{Mao},
extended sources \cite{Yoo}, and parallax effects \cite{Gould}
due to the non rectilinear apparent motion of the lens induced by the Earths' orbital motion.

\subsection{Statistical observables}
The optical depth up to a given source distance, $D_S$, is defined as the
probability to intercept
a deflector's Einstein disk, which corresponds to a magnification $A > 1.34$.
It is found to be independent of the deflectors' mass function
\begin{equation}
\tau=\frac{4 \pi G D_S^2}{c^2}\int_0^1 x(1-x)\rho(x) \ud x\, ,
\end{equation}
where $\rho(x)$ is the mass density of deflectors at distance $x D_S$.

Contrary to the optical depth, the microlensing event durations $t_E$ and consequently
the event rate (deduced from the optical depth and durations) depend on the deflectors' mass distribution
as well as on the velocity and spatial distributions.

The expected number of events for a microlensing survey is estimated from
\begin{equation}
N_{expect}= \frac{2}{\pi}\times \tau .N_{stars}T_{obs}  \int \frac{\epsilon(t_E)}{t_E}D(t_E)\ud t_E,
\label{event_number}
\end{equation}
where $N_{stars}$, $T_{obs}$, $\epsilon(t_E)$, and $D(t_E)$ are the
number of monitored stars, the survey duration, the detection efficiency, and the
normalized prior $t_E$ distribution of ongoing microlensing events at a given time.

\section{The four main surveys toward LMC and their results}
\label{section:surveys}
EROS, a mostly French collaboration, started to observe the LMC in the early 1990s. The first phase of this project consisted of two programs. The first program (\textit{EROS 1 plate}) used 290 digitized photographic plates taken at the ESO 1m Schmidt telescope from 1990 to 1994 \cite{eros1}. In the second program (1991-1994) a CCD camera was used to observe a small field in the LMC bar.
The fast cadence of this program made it sensitive to the low mass lenses with an event duration between 1 hour to 3 days \cite{eroslmc}. The observations towards the LMC continued in the second phase of EROS project (\textit{EROS 2}) using the Marly 1 meter telescope at ESO, La Silla, equipped with two 0.95 deg$^2$ CCD mosaics \cite{Tisserand2007}.
\par The MACHO team, a US-Australian collaboration, accumulated data on LMC from 1992 for nearly 6 years with a $1.27 m$ telescope and two cameras of 4 CCDs each, at the Australian Mount Stromlo site \cite{Alcock2000b}.
\par The OGLE experiment, a collaboration led by the University of Warsaw, started in 1992 and
is now in its fourth phase \cite{desc-OGLE4}. OGLE-III used a dedicated $1.3m$ telescope with a $64$ Mpixels camera.
Since 2009, OGLE-IV is equipped with a $256$ Mpixels large field camera.
\par The MOA team is a New Zealand and Japanese collaboration, operating a dedicated $1.8m$
telescope, equipped with a $64$ Mpixels camera \cite{MOAcam3}.
Since the published information on the MOA database towards the LMC is not as complete as the other collaborations,
we could not quantify a potential income to a combined analysis.
\par
\begin{figure}[htbp]
\begin{center}
\includegraphics[width=8cm]{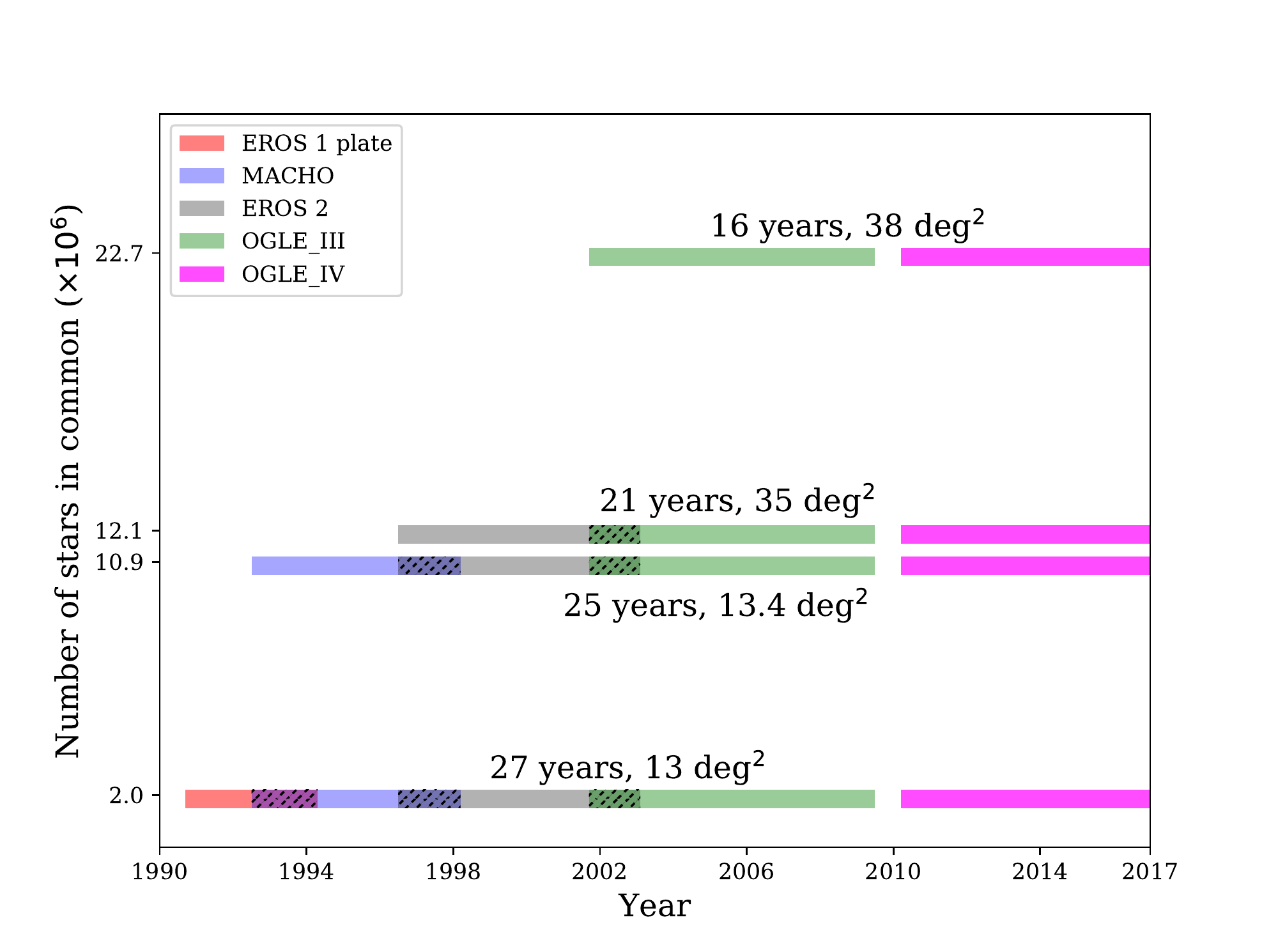}
\includegraphics[width=6cm]{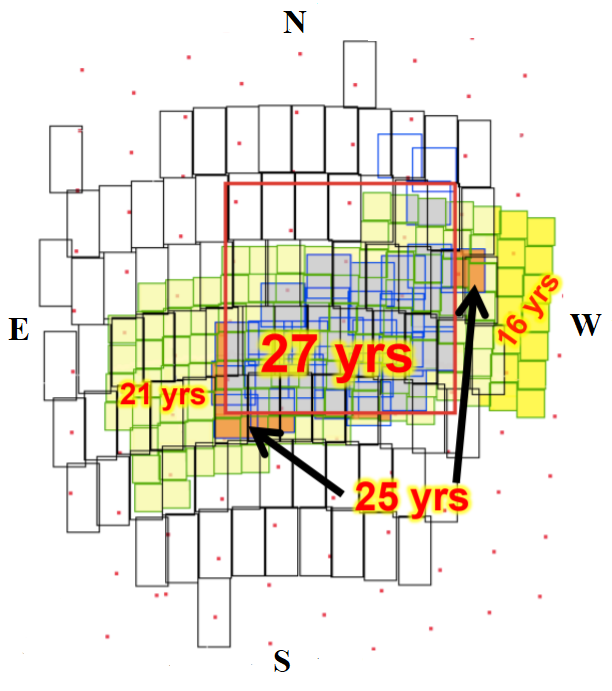}
\caption[]{(left) Catalogs of stars monitored by the different surveys of LMC; numbers of stars, field sizes, and observation epochs. Hatched regions show overlapping periods between different surveys. \\
-- (right) Fields monitored towards the LMC: EROS 1 plate (large red square), MACHO fields \cite{Alcock2000b} used to measure the optical depth (blue squares), EROS-2 (black rectangles), OGLE-III (green squares). Red dots show positions of (a fraction) of OGLE-IV field centers. The field colored in grey is observed for 27 years. The orange+grey field is observed for 25 years. The light yellow + orange + grey field is observed for 21 years. Finally, the union of all the colored fields is observed for 16 years.}
\label{timespan}
\end{center}
\end{figure}

\begin{figure}[htbp]
\begin{center}
\includegraphics[width=10cm]{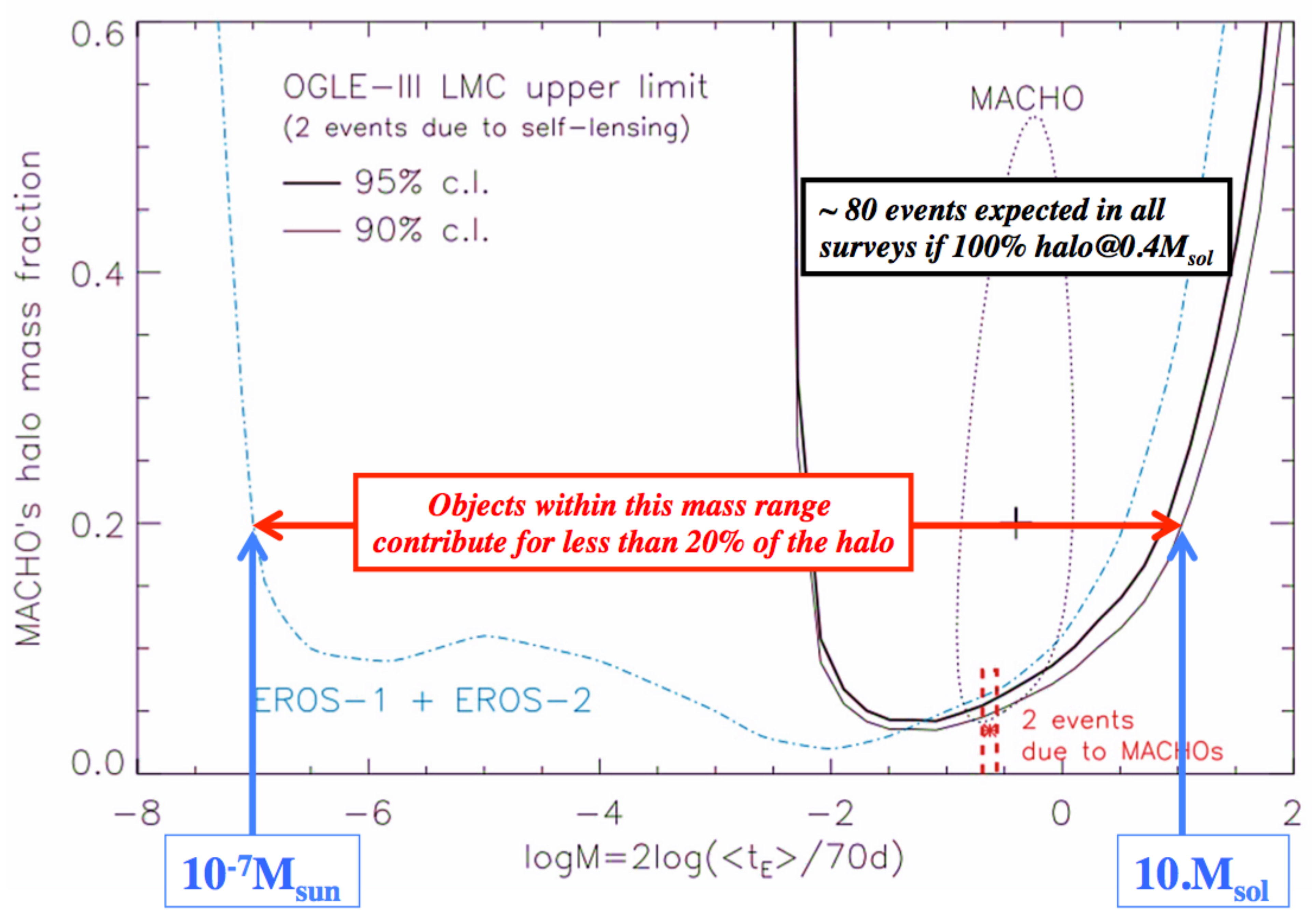}
\caption[]{
Constraints on the fraction of the standard spherical Galactic halo
made of massive compact objects as a function of their mass $M$.
The dotted-dashed line labeled EROS1+EROS2 shows the EROS $95\%$ upper
limit \cite{Tisserand2007}. The OGLE upper limit \cite{OGLE2010} is shown in full line.
The closed domain is the $95\%$ CL contour claimed by MACHO \cite{Alcock2000b}.
}
\label{resultatsLMC}
\end{center}
\end{figure}

 Table \ref{surveys} summarizes the EROS, MACHO, OGLE and MOA data taken towards the LMC. The positions of the fields monitored by these surveys towards the LMC (except the EROS 1 CCD program) are shown in the figure \ref{timespan} (right).
\begin{table}
{\scriptsize
\center
\begin{tabular}{|c|c|c|c|c|c|c|c|}
\hline
     Survey&Field (deg$^2$)&Filters&limit Mag.& Cadence &Objects ($\times 10^6$)&observation epochs&T$_{obs}$ (years)\\
     \hline
     \hline
     \multirow{2}{*}{EROS 1 plate}&\multirow{2}{*}{27.0}&R$_E$&\multirow{2}{*}{21.5}&\multirow{2}{*}{3 days}&\multirow{2}{*}{4.2}&\multirow{2}{*}{September 1990- April 1994 }&\multirow{2}{*}{3.5}\\
     &&B$_{E}$&&&&&\\
     \hline
     \multirow{2}{*}{MACHO}&\multirow{2}{*}{13.4}&R&\multirow{2}{*}{22}&\multirow{2}{*}{4 days}&\multirow{2}{*}{11.9}&\multirow{2}{*}{July 1992- March 1998}&\multirow{2}{*}{5.7}\\
     &&V&&&&&\\
     \hline
     \multirow{2}{*}{EROS 2}&\multirow{2}{*}{84.0}&I&\multirow{2}{*}{23}&\multirow{2}{*}{3 days}&\multirow{2}{*}{29.2}&\multirow{2}{*}{July 1996- February 2003}&\multirow{2}{*}{6.7}\\
     &&V&&&&&\\
     \hline
     \multirow{2}{*}{OGLE-III}&\multirow{2}{*}{38.0}&I&\multirow{2}{*}{23.4}&\multirow{2}{*}{3-4.6 days}&\multirow{2}{*}{22.7}&\multirow{2}{*}{September 2001- May 2009}&\multirow{2}{*}{7.7}\\
     &&V&&&&&\\
     \hline
     \multirow{2}{*}{OGLE-IV}&\multirow{2}{*}{$>$84}&I&\multirow{2}{*}{-}&\multirow{2}{*}{4 days}&\multirow{2}{*}{62}&\multirow{2}{*}{March 2010- present}&\multirow{2}{*}{-}\\
     &&V&&&&&\\
     \hline
    \multirow{3}{*}{MOA2-cam3}&\multirow{3}{*}{31.0}&I&\multirow{3}{*}{22.5}&\multirow{3}{*}{1 hour}&\multirow{3}{*}{50}&\multirow{3}{*}{2006-}&\multirow{2}{*}{-}\\
     &&R&&&&&\\
     &&V&&&&&\\
     \hline
\end{tabular}
}
\caption{The LMC field size, filters, maximum magnitude, cadence, number of monitored objects, observation epochs and duration for each survey.}
\label{surveys}
\end{table}
The main result from the LMC/SMC surveys \cite{Moniez} is that
compact objects of mass within $[10^{-7},10]\times \Msol$
interval are not a major component of the
hidden Galactic mass (Fig. \ref{resultatsLMC}).
We propose to explore the domain of the heavy compact objects (with $M>10.\Msol$),
that should produce long time-scale microlensing events, by combining the light-curves obtained
by all the surveys.
\section{Expectations from a simple combined analysis of LMC data}
\label{section:combined_analysis}
To look forward a combined analysis, we propose to use the full light-curve database of all the surveys. We have identified four star samples that have been monitored almost continuously for $\Delta T'=$ 16, 21, 25 and 27 years respectively, by at least two surveys towards the LMC (see Fig. \ref{timespan} (left)). The corresponding fields are shown in figure \ref{timespan} (right). Here below, we estimate the numbers of expected microlensing events from the populations of each sample and the corresponding extrapolated average detection efficiencies $\epsilon^{\Delta T'}(t_E)$ of a joint analysis with overall duration $\Delta T'$.

\subsection{Estimate of a combined analysis efficiency}
\label{section:efficiency}
\par It is well known that the efficiency of all surveys vanishes for long events due to their limited durations, but
we can conservatively extrapolate the global efficiency of a combined survey for much longer events from two basic hypotheses:
First, assuming a constant sampling rate, the detection efficiency of a survey increases for longer observations; second, the efficiency satisfies the following scaling invariance by time dilation by a factor $k$:
\begin{equation}
\epsilon(k\times t_E, LC(k\times t_i))=\epsilon(t_E, LC(t_i)) ,
\end{equation}
where $\epsilon(t_E, LC(t_i))$ is the efficiency to detect microlensing events with Einstein duration $t_E$ from a light-curve $LC$ defined by a series of flux measurements at times $t_i$, and $LC(k\times t_i)$ is the same series of flux measurements, but considered at times $k\times t_i$ instead of $t_i$.
Since the extended light-curve obtained by joining surveys should contain more measurements than the simply time-dilated light-curve,
these hypotheses allow us to conclude that the detection efficiency for longer events of a combined survey with overall duration $\Delta T'$ should be:
\begin{equation}
\epsilon^{\Delta T'}(t_E) \geq \epsilon^{\Delta T}(t_E\times \frac{\Delta T}{\Delta T'}),
\end{equation}
where $\Delta T$ and $\epsilon^{\Delta T}$ are the 
duration and efficiency of the longest individual survey.
We therefore define conservative extrapolated efficiencies as (see \cite{MEMO} for details): 
\begin{equation}
\epsilon_c^{\Delta T'}(t_E) = max\left( \epsilon^{\Delta T}(t_E)_{OGLE}\,,\,\epsilon^{\Delta T'}(t_E)_{OGLE}\right).
\end{equation}
\subsection{Number of expected microlensing events}
\label{section:results}
To estimate the number of expected events $N_{exp}$, we distinguish between the stellar populations monitored during $\Delta T'=$ 25, 21 and 16 years \footnote{The impact of the EROS1-plate survey was found negligible, due to the small size of its catalog and its
small marginal contribution to the overall time coverage (from 25 to 27 years).},
and consider the corresponding extrapolated efficiencies $\epsilon_c^{\Delta T'}(t_E)$. 
Each population number $N_{\Delta T'}$ has been estimated in its field (Fig. \ref{timespan} (right)) from the mean density associated to the shallowest survey:
\begin{equation}
    N_{exp} = \frac{2\tau}{\pi}\times
    \sum_{\Delta T'\, in\, \{16,21,25\}} \Delta T' N_{\Delta T'}\int \frac{\epsilon_c^{\Delta T'}(t_E)}{t_E} D(t_E)\ud t_E ,
\end{equation}
With this procedure, the expectation for each star is estimated from its most complete light-curve obtained by combining all available surveys.

\section{Discussion}
\label{section:discussion}
\begin{figure}[htbp]
\begin{center}
\includegraphics[width=9cm]{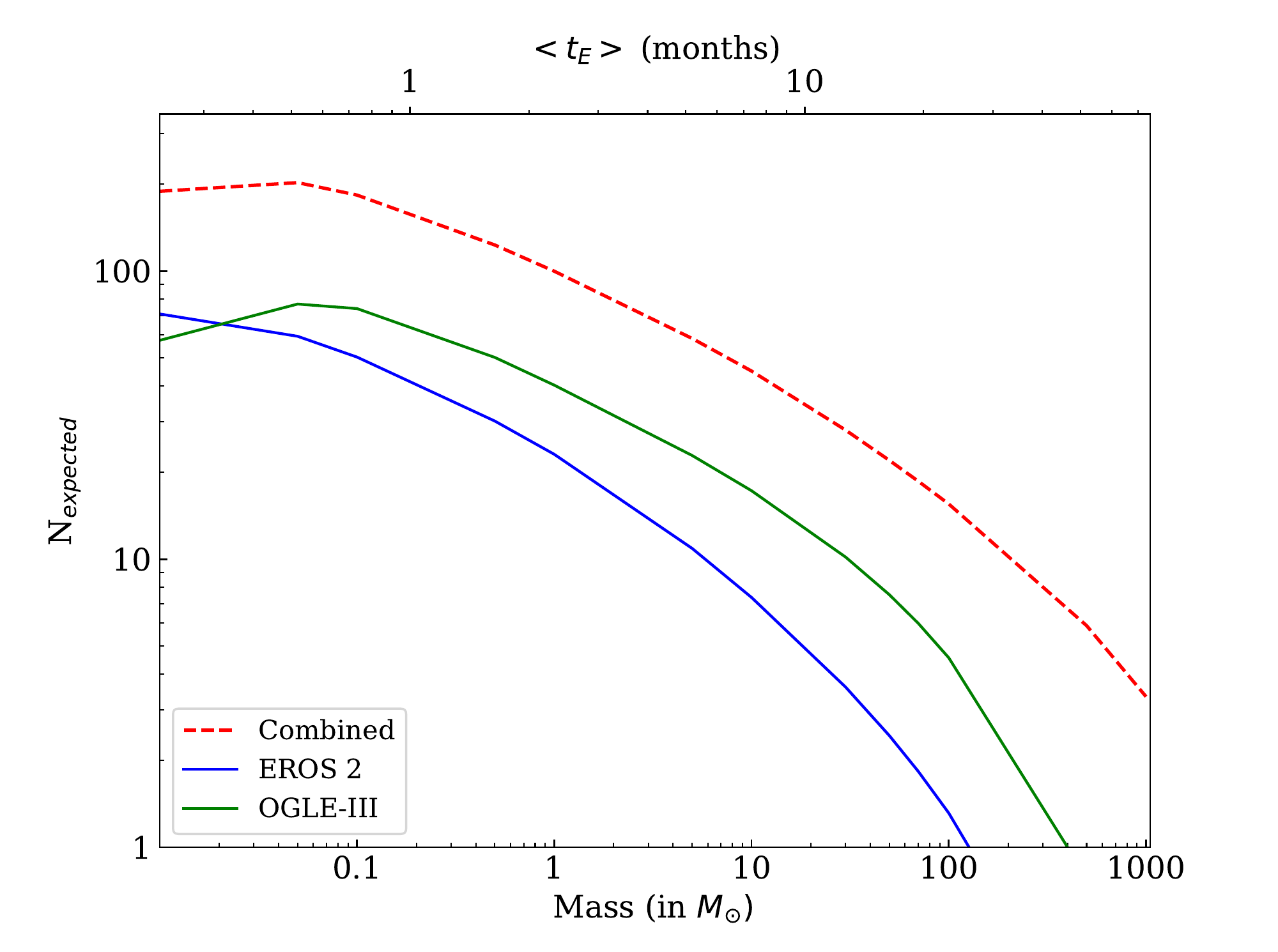}
\caption{Number of expected events for EROS 2 (blue) and OGLE-III (black) and for a combined analysis of the MACHO, EROS and OGLE-III+IV surveys (dashed-red line), assuming a Galactic halo made of mono-mass (lower abscissa) compact objects. The upper abscissa gives $<t_E>$ corresponding to the deflectors mass.
}
\label{expectation}
\end{center}
\end{figure}

The total number of expected events is shown in Fig. \ref{expectation}. The numbers of events expected for EROS 2 and OGLE-III surveys alone are shown for comparison.
If the Galactic halo consists in $100M_{\odot}$ objects, then EROS2 alone would have detected $\sim 1$ event and OGLE-III alone expects $\sim 5$ events, since $\sim 15$ events are expected by combining all the surveys.
This potential encourages us to perform such joint analysis.
Here, the extrapolated efficiencies assume that all events are point-source, point-lens, with rectilinear relative motion, an approximation
which is valid for $90\%$ ot the events.
The parallax effect significantly distorts the shape of the light-curves only if the Earth orbital velocity ($30 km/s$) is not negligible compared with the lens transverse velocity projected in the solar transverse plane ($\tilde{v} = v_T/(1-x)$).
Toward LMC, less than $\sim 5\%$ of the events due to standard halo lenses heavier than $10M_{\odot}$ are expected
to have $\tilde{v} < 5\times 30 km/s$ \cite{Rahvar2003}, and the light-curve distortion is almost never large enough to significantly affect the detection
efficiencies.

Among the expected complications, we also want to mention that the different colour passbands between the surveys, and the inter-seasonal
telescopes' throughput steps will also need a special care to limit misleading  preselections of long time-scale variations. In this purpose,
the colour equations between the different passbands and the zero-points of the instrumental magnitudes will have to be precisely established.
Then a precise global microlensing detection efficiency will need to be computed from specific simulations of combined light-curves, taking into account the proper photometric uncertainties and blending belonging to each survey.

\section{Conclusions and perspectives}
In this study, we showed that the joint analysis of the complete available data from all the past, present and future microlensing surveys towards the Large Magellanic Cloud, has the potential to at least double the long time-scale event rate expected by the most sensitive survey alone.
Several stages can be foreseen for such a combined analysis:
the first step -- corresponding to Fig. \ref{expectation} -- simply uses the already existing light-curve catalogs; a second step would
consist in the complete re-analysis of all the images, searching for variabilities through differential photometry.
Such a combined effort should allow one to estimate the black holes contribution to
the Milky-way halo and help to quantify the rate of gravitational wave detections.

\end{document}